\documentclass[12pt]{article}
\usepackage{a4wide,cite,graphicx,latexsym,amssymb}

\newcommand{\RE}{\mbox{\rm Re}}
\newcommand{\eqn}[1]{(\ref{#1})}
\newcommand{\mev}{\mbox{\rm MeV}}
\newcommand{\gev}{\mbox{\rm GeV}}

\begin{document}
%\bibliographystyle{../../aux/physics}

%%%%%%%%%%%%%%%%%%%%%%%%%%%%%%%%%%%%%%%%%%%%%%%%%%%%%%%%%%%%%%%%%%%%%%%%%
% The title page
%%%%%%%%%%%%%%%%%%%%%%%%%%%%%%%%%%%%%%%%%%%%%%%%%%%%%%%%%%%%%%%%%%%%%%%%%

\begin{titlepage}

\phantom{x}\vspace{-1cm}
% \phantom{x}\vspace{-1cm}
% \begin{center}
% {\Large Version of: 9.5.2006}
% \end{center}

\begin{flushright}
{UAB-FT-601}\\
{IFIC/06-10}\\
{FTUV/06-0509}\\[16mm]
\end{flushright}

\begin{center}
{\Large\sf\bf Spectral distribution for the decay
 \boldmath{$\tau\to\nu_\tau K\pi$}}\\[12mm]

{\large\bf Matthias Jamin${}^{1}$, Antonio Pich${}^{2}$, and}
{\large\bf Jorge Portol\'es${}^{2}$}\\[10mm]

{\small\sl ${}^{1}$ Instituci\'o Catalana de Recerca i Estudis
 Avan\c{c}ats (ICREA)} and\\
{\small\sl Departament de F\'{\i}sica Te\`orica, IFAE, UAB, E-08193 Bellaterra,
 Barcelona, Spain}\\[3mm]
{\small\sl ${}^{2}$ Departament de F\'{\i}sica Te\`orica, IFIC,
 Universitat de Val\`encia -- CSIC,}\\
{\small\sl Apartat de Correus 22085, E-46071 Val\`encia, Spain.}\\[2cm] 
\end{center}

{\bf Abstract:}
With the newly available data sets on hadronic $\tau$ decays from the
$B$-factories BABAR and BELLE, and future data from BESIII, precise information
on the decay distributions will soon become available. This calls for an
improvement of the decay spectra also on the theoretical side. In this work,
the distribution function for the decay $\tau\to\nu_\tau K\pi$ will be
presented with the relevant $K\pi$ vector and scalar form factors being
calculated in the framework of the resonance chiral theory, also taking into
account additional constraints from dispersion relations and short-distances.
As a by-product the slope and curvature of the vector form factor
$F_+^{K\pi}(s)$ are determined to be $\lambda_+^{'}=25.6\cdot 10^{-3}$ and
$\lambda_+^{''}=1.31\cdot 10^{-3}$ respectively. From our approach it appears
that it should be possible to obtain information on the low lying scalar
$K_0^*(800)$ as well as the second vector $K^*(1410)$ resonances from the
$\tau$ decay data. In particular, the exclusive branching fraction of the
scalar component is found to be
$B[\tau\to\nu_\tau(K\pi)_{\rm S-wave}]=(3.88\pm 0.19)\cdot 10^{-4}$.

\vfill


\end{titlepage}

\newpage
\setcounter{page}{1}

%%%%%%%%%%%%%%%%%%%%%%%%%%%%%%%%%%%%%%%%%%%%%%%%%%%%%%%%%%%%%%%%%%%%%%%%%
% The main part of the paper
%%%%%%%%%%%%%%%%%%%%%%%%%%%%%%%%%%%%%%%%%%%%%%%%%%%%%%%%%%%%%%%%%%%%%%%%%

\section{Introduction}

Already more than a decade ago it was realised that the hadronic decays of
the $\tau$ lepton could serve as an ideal system to study low-energy QCD under
rather clean conditions \cite{bnp92,bra89,bra88,np88,pich89}. In the following
years, detailed investigations of the $\tau$ hadronic width as well as
invariant mass distributions have allowed to determine many QCD parameters, a
most prominent example being the QCD coupling $\alpha_s$. Especially the
experimental separation of the Cabibbo-allowed decays and Cabibbo-suppressed
modes into strange particles \cite{dhz05,opal04,aleph99} opened a means to also
determine the quark-mixing matrix element $|V_{us}|$
\cite{gjpps04,jam03,gjpps03} as well as the mass of the strange quark
\cite{bck04,cdghpp01,dchpp01,km00,kkp00,pp99,ckp98,pp98}, additional
fundamental parameters within the Standard Model, from the $\tau$ strange
spectral function.

The dominant contribution to the Cabibbo-suppressed $\tau$ decay rate is due
to the decay $\tau\to\nu_\tau K\pi$. The corresponding distribution function
has been measured experimentally in the past by ALEPH \cite{aleph99} and OPAL
\cite{opal04}. With the large data sets on hadronic $\tau$ decays from the
B-factories BABAR and BELLE, which are currently under investigation, and good
prospects for additional data from BESIII in the future, a refined theoretical
understanding of the spectral functions is called for.  For the decay in
question, the general expression for the differential decay distribution takes
the form \cite{fm96}
\begin{equation}
\label{dGtau2kpi}
\frac{d\Gamma_{K\pi}}{d\sqrt{s}} \;=\; \frac{G_F^2|V_{us}|^2 M_\tau^3}{32\pi^3s}
\,\biggl(1-\frac{s}{M_\tau^2}\biggr)^2\Biggl[\, \biggl(1+2\,\frac{s}{M_\tau^2}
\biggr) q_{K\pi}^3\,|F_+^{K\pi}(s)|^2 + \frac{3\Delta_{K\pi}^2}{4s}\,q_{K\pi}
|F_0^{K\pi}(s)|^2 \,\Biggr] \,,
\end{equation}
where we have assumed isospin invariance and have summed over the two possible
decays $\tau^-\to\nu_\tau\overline K^0\pi^-$ and $\tau^-\to\nu_\tau K^-\pi^0$,
with the individual decay channels contributing in the ratio
$2\hspace{-0.4mm}:\!1$ respectively.  In this expression, $F_+^{K\pi}(s)$ and
$F_0^{K\pi}(s)$ are the vector and scalar $K\pi$ form factors respectively
which will be explicated in more detail below,
$\Delta_{K\pi}\equiv M_K^2-M_\pi^2$, and $q_{K\pi}$ is the Kaon momentum in the
rest frame of the hadronic system,
\begin{equation}
q_{K\pi}(s) \;=\; \frac{1}{2\sqrt{s}}\sqrt{\Big(s-(M_K+M_\pi)^2\Big)
\Big(s-(M_K-M_\pi)^2\Big)}\cdot\theta\Big(s-(M_K+M_\pi)^2\Big) \,.
\end{equation}

By far the dominant contribution to the decay distribution originates from the
$K^*(892)$ meson. In the next section, an effective description of this
contribution to the vector form factor $F_+^{K\pi}(s)$ will be presented in
the framework of chiral perturbation theory with resonances (R$\chi$PT)
\cite{egpr89,eglpr89}, quite analogously to a similar description of the
pion form factor given in refs.~\cite{gp97,pp01,scp02}. To also include a
second vector resonance, the effective chiral description can be
straightforwardly augmented by an additional $K^*(1410)$ meson. Finally, the
scalar $K\pi$ form factor $F_0^{K\pi}(s)$ has very recently been updated in
ref.~\cite{jop06}. In section~3, we shall present our main results for the
distribution function $d\Gamma_{K\pi}/d\sqrt{s}$ and total decay rates and
investigate the influence of the different vector and scalar form factor
contributions.

\section{The vector form factor \boldmath{$F_+^{K\pi}(s)$}}

A theoretical representation of the vector form factor $F_+^{K\pi}(s)$, which
is based on fundamental principles, can be provided by an effective field
theory description, in complete analogy to the description of the pion form
factor presented in refs.~\cite{gp97,pp01,scp02}\footnote{For an alternative
dispersive approach to the pion form factor see also refs.~\cite{mn02,mnnp04}.}.
This approach employs our present knowledge on effective hadronic theories,
short-distance QCD, the large-$N_C$ expansion as well as analyticity and
unitarity. For the pion form factor the resulting expressions provide a very
good description of the experimental data \cite{gp97,pp01,scp02}.

Close to $s$ equal zero, $F_+^{K\pi}(s)$ is well described by the $\chi$PT
result at one loop \cite{gl85b}, which takes the form:
\begin{equation}
\label{FpChPT}
F_+^{K\pi}(s) \;=\; 1 + \frac{2}{F_\pi^2}\,L_9^r\,s + \frac{3}{2}\Big[
\widetilde{H}_{K\pi}(s) + \widetilde{H}_{K\eta}(s) \Big] \,.
\end{equation}
The one-loop function $\widetilde{H}(s)$ is related to the corresponding
function $H(s)$ of \cite{gl85b} by
$\widetilde{H}(s)\equiv H(s) - 2L_9^r\,s/(3F_\pi^2)=[s M^r(s)-L(s)]/F_\pi^2$.
Explicit expressions for $M^r(s)$ and $L(s)$ can be found in ref.~\cite{gl85}.
The vector form factor is an analytic function in the complex $s$ plane,
except for a cut along the positive real axis, starting at
$s_{K\pi}\equiv (M_K+M_\pi)^2$, where its imaginary part develops a
discontinuity. In the elastic region below roughly $1\,\gev$, $F_+^{K\pi}(s)$
admits the well-known Omn\`es representation \cite{om58}
\begin{equation}
\label{omel}
F_+^{K\pi}(s)\;=\;P(s) \exp \Biggl( \frac{s}{\pi} \int\limits^\infty_{s_{K\pi}}
\!\!ds'\, \frac{\delta_1^{1/2}(s')}{s'(s'-s-i0)}\Biggr) \,,
\end{equation}
with $P(s)$ being a real polynomial to take care of the zeros of $F_+^{K\pi}(s)$
for finite $s$ and $\delta_1^{1/2}(s)$ is the P-wave $I=1/2$ elastic $K\pi$
phase shift.

Precisely following the approach of ref.~\cite{gp97}, and matching the Omn\`es
formula \eqn{omel} with the $\chi$PT calculation of $F_+^{K\pi}(s)$ in the
presence of vector resonances \cite{egpr89}, one finds the following
representation of the form factor $F_+^{K\pi}(s)$:
\begin{equation}
\label{FpEFT}
F_+^{K\pi}(s) \;=\; \frac{M_{K^*}^2 {\rm e}^{{\frac{3}{2}\RE\big[
\widetilde{H}_{K\pi}(s)+\widetilde{H}_{K\eta}(s)\big]}}}{M_{K^*}^2 -
s - iM_{K^*}\Gamma_{K^*}(s)} \;.
\end{equation}
The one-loop function $\widetilde{H}(s)$ depends on the chiral scale $\mu$,
and in eq.~\eqn{FpEFT}, this scale should be taken as $\mu=M_{K^*}$. In ref.
\cite{GomezDumm:2000fz}, the off-shell width of a vector resonance was defined
through the two-point vector current correlator, performing a Dyson-Schwinger
resummation within R$\chi$PT \cite{egpr89,eglpr89}. Following this scheme the
energy-dependent width $\Gamma_{K^*}(s)$ is found to be
\begin{equation}
\label{GaKs}
\Gamma_{K^*}(s) \;=\; \frac{G_V^2 M_{K^*} s}{64\pi F_\pi^4} \Big[
\sigma_{K\pi}^3(s) + \sigma_{K\eta}^3(s) \Big] \,,
\end{equation}
where $G_V$ is the chiral vector coupling which appears in the framework of
$\chi$PT with resonances \cite{egpr89}, the phase space function
$\sigma_{K\pi}(s)$ is given by $\sigma_{K\pi}(s)=2q_{K\pi}(s)/\sqrt{s}$, 
and $\sigma_{K\eta}(s)$ follows analogously with the replacement
$M_\pi\to M_\eta$. Re-expanding eq.~\eqn{FpEFT} in $s$ and comparing with
eq.~\eqn{FpChPT}, one reproduces the short-distance constraint for the vector
coupling $G_V=F_\pi/\sqrt{2}$ \cite{eglpr89} which guarantees a vanishing form
factor at $s$ to infinity, as well as the lowest-resonance estimate\footnote{A
recent one-loop calculation of the pion form factor in R$\chi$PT \cite{crp04}
has found that the NLO corrections to this lowest-resonance approximation are
very small.} for the ${\cal O}(p^4)$ chiral coupling \cite{egpr89}
\begin{equation}
L_9^r(M_{K^*}) \;=\; \frac{F_\pi^2}{2M_{K^*}^2} = 5.34 \cdot 10^{-3} \,,
\end{equation}
where $F_\pi=92.4\,\mev$ and the average mass of the charged and neutral
$K^*(892)$, $M_{K^*} = 893.9\,\mev$ have been used. This result is in very
good agreement to a recent determination of $L_9^r$ from the pion form factor
\cite{bt02} which found $L_9^r(M_{K^*}) = (5.69\pm 0.41)\cdot 10^{-3}$.

\begin{figure}[thb]
\begin{center}
\includegraphics[angle=0, width=15cm]{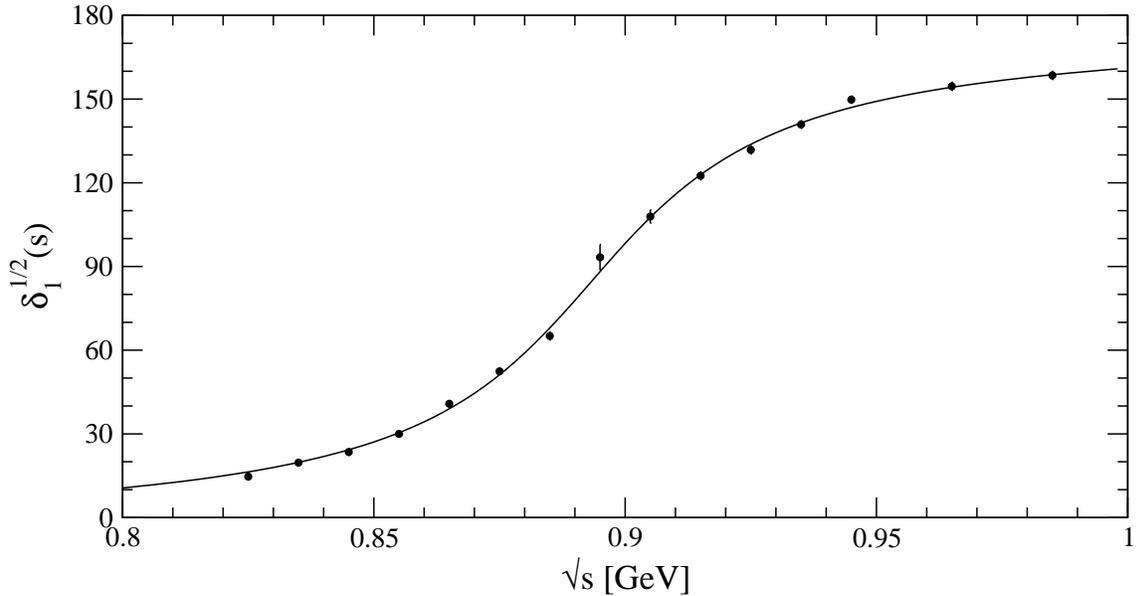}
\end{center}
\caption{$K\pi$ scattering P-wave phase-shift data of ref.~\cite{ast88},
together with our fit described in the text.\label{fig1}}
\end{figure}

From eq.~\eqn{FpEFT}, one obtains a description of the P-wave $K\pi$ phase
shift $\delta_1^{1/2}(s)$:
\begin{equation}
\label{del1}
\delta_1^{1/2}(s) \;=\; {\rm arctan}\biggl(\frac{M_{K^*}\Gamma_{K^*}(s)}
{M_{K^*}^2 - s}\biggr) \,.
\end{equation}
Improving the $K^*$-meson width $\Gamma_{K^*}(s)$ with the pertinent
Blatt-Weisskopf barrier factor $D_1(r q_{K\pi}(M_{K^*}^2))/D_1(r q_{K\pi}(s))$
\cite{bw52}, where $D_1(x)=1+x^2$ and $r$ is the interaction radius, we can
perform a fit of eq.~\eqn{del1} to the phase-shift data of
Aston~et~al.~\cite{ast88}. The fit parameters $M_{K^*}$, $G_V$ and $r$ are
found to be
\begin{equation}
\label{fit}
M_{K^*} \;=\; 895.9 \pm 0.5 \,\mev \,, \quad
G_V \;=\; 64.6 \pm 0.4 \,\mev \,, \quad
r \,=\, 3.5 \pm 0.6 \,\gev^{-1} \,,
\end{equation}
where the given errors are purely statistical and the corresponding fit is
displayed
in figure~\ref{fig1}.

We observe that the obtained $K^*$ mass is in complete agreement with the fit
result of ref.~\cite{ast88}, and also with the PDG average for the mass of the
neutral $K^*$ meson $M_{K^*} = 896.1 \pm 0.3\,\mev$ \cite{pdg04}, the relevant
channel in this case. Our result for $G_V$ is nicely compatible with the
short-distance constraint $G_V=F_\pi/\sqrt{2}\approx 65.3\,\mev$ found in
ref.~\cite{eglpr89}. Furthermore, our fit result for $r$ and 
our $\chi^2/$NDF$\;\approx 18.5/12$ are similar to
the ones obtained by Aston et al. \cite{ast88}. 
However, the result for $G_V$ found above
would correspond to a $K^*$ width
\begin{equation}
\Gamma_{K^*} \;\equiv\; \Gamma_{K^*}(M_{K^*}^2) \;=\; 55.8 \pm 0.8 \,\mev \,,
\end{equation}
which is about 10\% larger than the finding of \cite{ast88} and the PDG
average $\Gamma_{K^*}=50.7 \pm 0.6 \,\mev$ \cite{pdg04}. At the origin of this
discrepancy may lie the role played by a heavier vector resonance, namely the
$K^*(1410)$, on which we shall comment further below. As concluded in
ref.~\cite{scp02}, $G_V$ goes down a few percent when a second multiplet is
included into the analysis of the elastic $\pi\pi$ scattering amplitude,
providing $G_V = 61.9 \pm 1.5\,\mev$ \cite{scp02} which would result in
$\Gamma_{K^*} = 51.2 \pm 0.7\,\mev$, in excellent agreement with the PDG
average.

Next, we can compare the low-energy expansion of the vector form factor
\eqn{FpEFT} with recent experimental measurements of its slope and curvature.
Let us define a general expansion of $F_+^{K\pi}(s)$ as
\begin{equation}
F_+^{K\pi}(s) \equiv F_+^{K\pi}(0)\Biggl[\, 1 + \lambda_+^{'}\frac{s}{M_\pi^2}
+ \frac{1}{2}\,\lambda_+^{''}\frac{s^2}{M_\pi^4} + \frac{1}{6}\,\lambda_+^{'''}
\frac{s^3}{M_\pi^6} + \ldots \,\Biggr] \,,
\end{equation}
where $\lambda_+^{'}$, $\lambda_+^{''}$ and $\lambda_+^{'''}$ are the slope,
curvature and cubic expansion parameter respectively. Numerically, when
employing the following isospin averages for the meson masses
$M_\pi=138.0\,\mev$, $M_K=495.7\,\mev$ and the mass of the charged $K^*$ meson,
$M_{K^*}=891.66\,\mev$ \cite{pdg04}, which is relevant for the leptonic decays
of the $K$ meson, these are found to be:
\begin{equation}
\label{lambdap}
\lambda_+^{'}   \;=\; 25.6\cdot 10^{-3} \,, \qquad
\lambda_+^{''}  \;=\; 1.31\cdot 10^{-3} \,, \qquad
\lambda_+^{'''} \;=\; 9.74\cdot 10^{-5} \,.
\end{equation}
The parametric uncertainties on these results are rather small. However, it is
difficult to estimate the systematic uncertainties and therefore we have chosen
not to attach errors to the results of eq.~\eqn{lambdap}. Recent experimental
results on the slope $\lambda_+^{'}$ and the curvature $\lambda_+^{''}$ have
been collected in table~\ref{tab:1}. One observes that our results of
eq.~\eqn{lambdap} are in nice agreement with the very recent measurement by
KLOE~\cite{kloe06}, and in reasonable agreement with ISTRA \cite{istra04} as
well as NA48~\cite{na4804}, however about $2.5\,\sigma$ away from the KTEV
result \cite{ktev04}.

\begin{table}[thb]
\begin{center}
\begin{tabular}{lcc}
\hline
Collaboration & $\lambda_+^{'}\,[10^{-3}]$ & $\lambda_+^{''}\,[10^{-3}]$ \\
\hline
ISTRA 04 \cite{istra04} & $\qquad 23.2 \pm 1.6\qquad$ & $\quad 0.84 \pm 0.41\quad$\\
KTEV 04 \cite{ktev04} & $20.64 \pm 1.75$ & $3.20 \pm 0.69$ \\
NA48 04 \cite{na4804} & $28.0 \pm 2.4$ & $0.2 \pm 0.5$ \\
KLOE 06 \cite{kloe06} & $25.5 \pm 1.8$ & $1.4 \pm 0.8$ \\
\hline
\end{tabular}
\end{center}
\caption{Recent experimental results on the slope $\lambda_+^{'}$ and the
curvature $\lambda_+^{''}$ of the vector form factor $F_+^{K\pi}(s)$ in units
of $10^{-3}$.\label{tab:1}}
\end{table}

Since the $\tau$ lepton can also decay hadronically into the second vector
resonance $K^{*'}\equiv K^*(1410)$, this particle should be included in our
parametrisation of the vector form factor $F_+^{K\pi}(s)$. A parametrisation
which is motivated by the R$\chi$PT framework \cite{egpr89,eglpr89} can be
written as follows:
\begin{equation}
\label{FpKsKsp}
F_+^{K\pi}(s) \,=\, \Biggl[\, \frac{M_{K^*}^2+\gamma\,s}{M_{K^*}^2 - s -
iM_{K^*}\Gamma_{K^*}(s)} - \frac{\gamma\,s}{M_{K^{*'}}^2 - s -
iM_{K^{*'}}\Gamma_{K^{*'}}(s)} \,\Biggr] {\rm e}^{{\frac{3}{2}
\RE\big[\widetilde{H}_{K\pi}(s)+\widetilde{H}_{K\eta}(s)\big]}}  \,.
\end{equation}
The relation of the parameter $\gamma$ to the R$\chi$PT couplings takes the
form $\gamma = F_V G_V/F_\pi^2 - 1$, when one assumes a vanishing form factor
at large $s$ in the $N_C$ to infinity limit. It is difficult, to asses a
precise value for $\gamma$, but from the work of ref.~\cite{scp02}, we infer
that it should be small and positive ($0<\gamma\ll 1$). We shall come back to
the parameter $\gamma$ below. The width of the second resonance cannot be set
unambiguously. Therefore, we have decided to endow the $K^*(1410)$ contribution
with a generic width as expected for a vector resonance. Hence,
$\Gamma_{K^{*'}}(s)$ will be taken to have the form
\begin{equation}
\Gamma_{K^{*'}}(s) \;=\; \Gamma_{K^{*'}}\,\frac{s}{M_{K^{*'}}^2}\Biggl(
\frac{\sigma_{K\pi}^3(s)}{\sigma_{K\pi}^3(M_{K^{*'}}^2)}\Biggr) \,.
\end{equation}

\section{The differential decay distribution}

As a last ingredient for a prediction of the differential decay distribution
of the decay $\tau\to\nu_\tau K\pi$ according to eq.~\eqn{dGtau2kpi}, we
require the scalar form factor $F_0^{K\pi}(s)$. This form factor was calculated
in a series of articles \cite{jop00,jop01a,jop01b} in the framework of
R$\chi$PT, again also employing constraints from dispersion theory as well as
the short-distance behaviour, which then lead to a determination of the strange
quark mass $m_s$ in \cite{jop01b}. Quite recently, the determination of
$F_0^{K\pi}(s)$ was updated in \cite{jop06} by employing novel experimental
constraints on the form factor at the Callan-Treiman point $\Delta_{K\pi}$,
and here we shall also make use of this update.

A remaining question is which value to use for the form factors $F_+^{K\pi}(s)$
and $F_0^{K\pi}(s)$ at the origin. From our chiral description of the vector
form factor \eqn{FpEFT}, one would obtain $F_+^{K\pi}(0) = 0.978$. An average
over recent determinations from lattice QCD and effective field theory
approaches \cite{jop04,bec04,cir05,daw05,tsu05,oka05}, however, yields
\begin{equation}
\label{Fp00}
F_+^{K\pi}(0) \;=\; F_0^{K\pi}(0) \;=\; 0.972 \pm 0.012 \,,
\end{equation}
somewhat lower and also compatible with the original estimate by Leutwyler
and Roos \cite{lr84}. Furthermore, this is the value to which the scalar form
factor had been normalised in ref.~\cite{jop06}. Nevertheless, what is required
in the hadronic $\tau$ decays is only the product $|V_{us}| F_+^{K\pi}(0)$ which
experimentally is known more precisely. Thus, in what follows, to normalise the
form factors, we shall employ the most recent average for this number
\cite{vus05}:
\begin{equation}
\label{VusF0}
|V_{us}| F_+^{K\pi}(0) \;=\; 0.2173 \pm 0.0008 \,.
\end{equation}

\begin{figure}[thb]
\begin{center}
\includegraphics[angle=0, width=15cm]{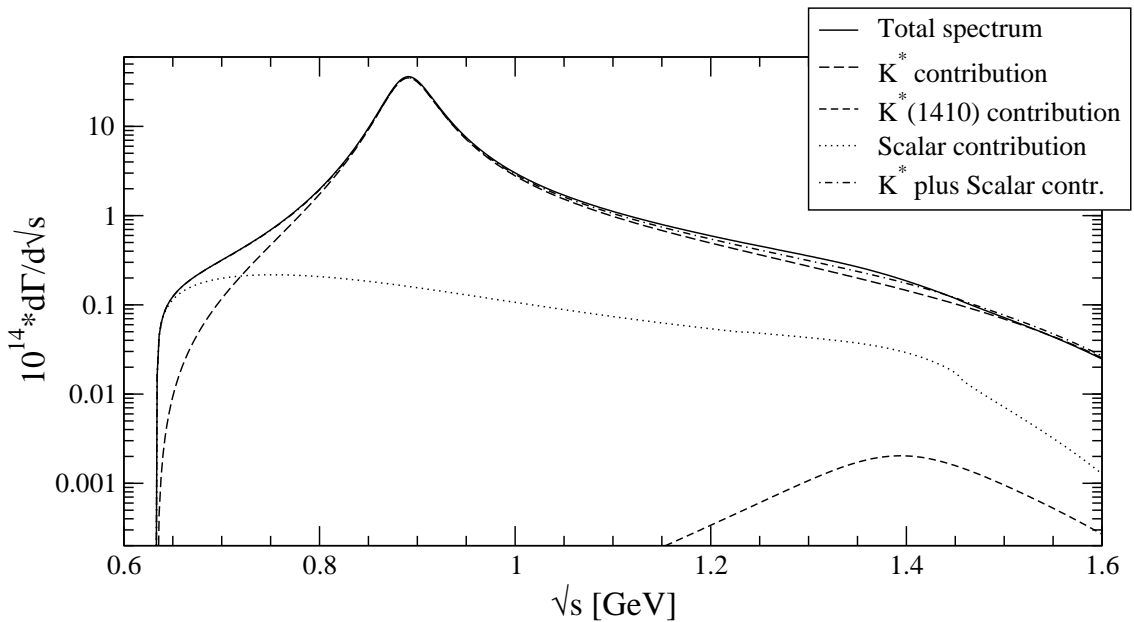}
\end{center}
\caption{Main result for the differential decay distribution of the decay
$\tau\to\nu_\tau K\pi$, together with the individual contributions from the
$K^*$ and $K^*(1410)$ vector mesons as well as the scalar component residing
in the scalar form factor $F_0^{K\pi}(s)$.\label{fig2}}
\end{figure}

Our main result for the differential decay distribution is displayed as the
solid line in figure~\ref{fig2}, together with the individual contributions.
Let us discuss our inputs and the individual contributions in more detail.
Since the $K^*$ meson in $\tau$ decays is the charged one, in contrast to the
fit result of eq.~\eqn{fit}, for the $K^*$ mass we have employed the PDG value
$M_{K^{*-}} = 891.66\pm 0.26\,\mev$ \cite{pdg04}. However, for $G_V$ and
the barrier factor parameter $r$, the fit results of \eqn{fit} have been used.
The resulting contribution of the $K^*$ meson to the spectral distribution is
shown as the long-dashed line in figure~\ref{fig2}. Integrating this part over
the phase space and varying the input parameters, one finds
\begin{equation}
B[\tau\to\nu_\tau K^*(892)] \;=\; (1.253^{+\,0.062}_{-\,0.076} \pm 0.019)\%
\;=\; (1.253 \pm 0.078)\% \,.
\end{equation}
The first uncertainty represents an estimate of higher order chiral corrections.
To this end, in the exponential of eq.~\eqn{FpEFT}, we have replaced the factor
$1/F_\pi^2$ by $1/(F_K F_\pi)$ or by $1/F_0^2$ with $F_0 = 87\,\mev$ being the
pion decay constant at the leading order, which should give an idea about
unaccounted further chiral corrections. The remaining uncertainty arises from a
variation of the fit parameters $G_V$ and $r$ of eq.~\eqn{fit} and the value
\eqn{VusF0} for $|V_{us}| F_+^{K\pi}(0)$.

Next, the contribution from the scalar form factor $F_0^{K\pi}(s)$ in figure
\ref{fig2} is displayed as the dotted line. Its most important contribution
arises in the region below the $K^*$ resonance where the low-lying scalar
$K_0^*(800)$ resonance is active. Integrating over the scalar contribution,
we obtain
\begin{equation}
\label{Bswave}
B[\tau\to\nu_\tau(K\pi)_{\rm S-wave}] \;=\; (3.88\pm 0.19)\cdot 10^{-4} \,,
\end{equation}
where the error dominantly is due to a variation of the form factor shape as
discussed in refs.~\cite{jop01a,jop04,jop06}. Since the scalar resonances are
not well described by Breit-Wigners and there is also a strong interference
between the dynamically generated $K_0^*(800)$ and the pre-existing (at
$N_C\to\infty$) $K_0^*(1430)$ resonance, we prefer not to resolve the $K\pi$
S-wave contribution into individual components. (For some remarks on the
$K_0^*(800)$, also known as the $\kappa$, see section~7 of ref.~\cite{jop00}.)
The sum of the scalar and $K^*$ contributions in figure~\ref{fig2} is shown as
the dashed-dotted line.

The last remaining contribution is the one due to the second vector resonance
$K^*(1410)$. For its mass and width, we have employed the PDG values
$M_{K^*(1410)} = 1414\,\mev$ and $\Gamma_{K^*(1410)} = 232\,\mev$ \cite{pdg04}.
The $K^*(1410)$ contribution turns out to depend very sensitively on the mixing
parameter $\gamma$ defined in eq.~\eqn{FpKsKsp}, for which an estimate can be
obtained on the basis of the total branching fraction $B[\tau\to\nu_\tau K\pi]$.
Adding the results of the most recent compilation \cite{dhz05}, one finds
$B[\tau\to\nu_\tau K\pi]=(1.33 \pm 0.05)\,\%$. Then adjusting $\gamma$ such
that the experimental total branching fraction, including its uncertainty, is
reproduced, we obtain $\gamma = 0.013 \pm 0.017$, in agreement with the
expectation of the last section, that $\gamma$ should be small and positive.
The contribution of the $K^*(1410)$ resonance with the central value of $\gamma$
is shown as the short-dashed line in figure~\ref{fig2}. Even though this
contribution appears rather small, because of the interference with the leading
$K^*$ resonance, its influence in the energy region above the $K^*$ resonance
is roughly as important as the scalar component. This is also reflected in the
corresponding total branching fraction which turns out to be
\begin{equation}
\label{BKs1410}
B[\tau\to\nu_\tau (K^*(892)+K^*(1410))] - B[\tau\to\nu_\tau K^*(892)]
\;\approx\; 3.9 \cdot 10^{-4} \,,
\end{equation}
however, with rather large uncertainties, but much bigger than the $K^*(1410)$
branching ratio by itself whose central value reads
$B[\tau\to\nu_\tau K^*(1410)]=2.1\cdot 10^{-6}$. We notice that the central
result \eqn{BKs1410} is similar to the scalar branching fraction \eqn{Bswave}
and compatible to an estimate presented by the ALEPH collaboration
\cite{aleph99} which yielded
$B[\tau\to\nu_\tau K^*(1410)]=(1.5^{+1.4}_{-1.0})\cdot 10^{-3}$, but where,
however, the scalar component had been neglected.

\section{Conclusions}

Upcoming much improved results on the branching fractions and differential
distributions of hadronic $\tau$ decays from the $B$-factories BABAR and
BELLE, as well as in the near future from BESIII, necessitate an analogous
improvement also on the theoretical side. A step in this direction is taken in
the work at hand, where we have presented a description of the decay spectrum
of the decay $\tau\to\nu_\tau K\pi$ \eqn{dGtau2kpi} in the framework of
R$\chi$PT \cite{egpr89,eglpr89}.

Our approach follows the lines of an analogous description for the pion form
factor \cite{gp97,pp01,scp02}, which employs all present knowledge of effective
theories, short-distance QCD, the large-$N_C$ expansion as well as analyticity
and unitarity, and was successful in describing the experimental data for the
pion form factor. Our central result for the $K\pi$ vector form factor, also
including the second vector resonance $K^*(1410)$ in this channel, has been
presented in eq.~\eqn{FpKsKsp}. As a by-product, we have determined the slope
and curvature of $F_+^{K\pi}(s)$,
\begin{equation}
\lambda_+^{'}   \;=\; 25.6\cdot 10^{-3} \,, \qquad\qquad
\lambda_+^{''}  \;=\; 1.31\cdot 10^{-3} \,,
\end{equation}
in very good agreement with the recent KLOE results \cite{kloe06}. The required
scalar $K\pi$ form factor $F_0^{K\pi}(s)$ has been employed from the recent
update \cite{jop06}, following the previous analyses \cite{jop00,jop01a}.

Our central result for the differential decay distribution
$d\Gamma_{K\pi}/d\sqrt{s}$ of eq.~\eqn{dGtau2kpi} is displayed in
figure~\ref{fig2}, where also the separate contributions originating from the
$K^*$, the $K^*(1410)$, and the scalar component have been shown
separately.\footnote{A Fortran routine to generate this distribution can be
obtained on request from the authors.} As can be observed from this figure,
the S-wave $K\pi$ contribution is most prominent in the energy region below
the $K^*$ resonance, and dominated by the dynamically generated $K_0^*(800)$
resonance (also known as the $\kappa$). Thus, by analysing experimental data
in this energy range, valuable information about the still much debated
$K_0^*(800)$ resonance can be obtained. Integrating the scalar component over
the phase space, we obtain
\begin{equation}
B[\tau\to\nu_\tau(K\pi)_{\rm S-wave}] \;=\; (3.88\pm 0.19)\cdot 10^{-4} \,,
\end{equation}
which is already rather precise, so that it will be difficult to improve this
accuracy experimentally.

The contribution of the second vector resonance $K^*(1410)$ dominantly depends
on the mixing parameter $\gamma$. However, this parameter can be inferred from
the total $\tau\to\nu_\tau K\pi$ branching fraction, resulting in the estimate
$\gamma = 0.013 \pm 0.017$. Even though the $K^*(1410)$ contribution by
itself is small, through interference with the $K^*$ resonance, in the range
around $1.4\,\gev$ it becomes about as important as the scalar resonance in this
region, the $K_0^*(1430)$. Therefore, with a dedicated experimental analysis
in the region above the $K^*$ peak, it should be possible to also obtain
information on the $K^*(1410)$.

In summary, in this work we have started a dedicated effort to improve the
description of exclusive strangeness-changing hadronic $\tau$-decays. We are
convinced that our description of the decay spectrum for the decay
$\tau\to\nu_\tau K\pi$ will proof valuable in the analysis of the
high-statistics data on hadronic $\tau$ decays, acquired at the $B$-factories
BABAR and BELLE. We plan to return to this subject with other decay channels
in the future.

\vskip 1cm \noindent
{\Large\bf Acknowledgements}

\noindent
M.J. would like to thank the IFIC, Valencia for great hospitality during a stay
in which part of this work has been completed. This work has been supported in
 part by the European Union (EU) RTN Network EURIDICE Grant
No. HPRN-CT2002-00311 (M.J., A.P. and J.P.), by MEC (Spain) and FEDER (EU)
Grants No. FPA2005-02211 (M.J.) and No. FPA2004-00996 (A.P. and J.P.), and
by Generalitat Valenciana Grants GRUPOS03/013, GV04B-594 and GV05/015
(A.P. and J.P.).

%\bibliography{tau2kpi}

\end{document}